\documentclass[twocolumn,5p,times]{article}
\usepackage{graphicx}
\usepackage{lscape,graphicx}
\usepackage{hhline}
\usepackage{multicol}
\usepackage{colordvi}
\usepackage{dcolumn}
\usepackage[fleqn]{amsmath}
\usepackage[mathscr]{euscript}
\usepackage[usenames, dvipsnames]{color}
\usepackage[usenames,dvipsnames,svgnames,table]{xcolor}
\usepackage[linktocpage]{hyperref}
\usepackage[a4paper,bindingoffset=0.2in,left=0.5in,right=0.5in,top=1.0in,bottom=1.0in,footskip=0.25in]{geometry}
\newcommand*{\doi}[1]{\href{http://dx.doi.org/#1}{doi: #1}}
\usepackage[numbers]{natbib}
\makeatletter
\renewcommand{\section}{\@startsection{section}{1}{\z@}{-3.5ex \@plus -1ex \@minus -.2ex}{1.3ex \@plus.2ex}{\normalfont\small\bfseries\boldmath}}
\makeatother

\makeatletter
\renewcommand{\subsection}{\@startsection{subsection}{2}{\z@}{-3.5ex \@plus -1ex \@minus -.2ex}{1.3ex \@plus.2ex}{\normalfont\small\bfseries\boldmath}}
\makeatother

\makeatletter
\renewcommand{\subsubsection}{\@startsection{subsubsection}{3}{\z@}{-3.5ex \@plus -1ex \@minus -.2ex}{1.3ex \@plus.2ex}{\normalfont\small\bfseries\boldmath}}
\makeatother

\makeatletter
\everymath{\if@display\else\thickmuskip=0mu plus 0mu\fi}
\makeatother

\title{\large \bf  {\tt ee$\in$MC}: Higher Orders of Radiative Emissions for $\bf e^{+}e^{-} \to \mu^{+}\mu^{-} (n\gamma)$  and $\bf e^{+}e^{-} \to \tau^{+}\tau^{-} (n\gamma)$ Events }
\date{}

\author{\normalsize Ian M. Nugent$^{*}$ \\ \normalsize Victoria, B.C., Canada}

\begin{document}
\twocolumn[
  \begin{@twocolumnfalse}
    \maketitle
\begin{abstract}
The radiative emissions in the {\tt ee$\in$MC} generator are extended to include the Initial/Final-State radiative diagrams, with Initial-State/Final-State interference, for the NLO and 
NNLO calculations and the Initial-State radiative diagrams associated with the NNNLO calculation for $e^{+}e^{-} \to \mu^{+}\mu^{-} (n\gamma)$  and $e^{+}e^{-} \to \tau^{+}\tau^{-} (n\gamma)$ Events. 
The NNNLO Initial-State radiative diagrams represent the largest contribution to the perturbative Feynman series in the emission of 4 hard-photons. From the convergence of the total cross-section
for $e^{+}e^{-} \to \mu^{+}\mu^{-} (n\gamma)$  and $e^{+}e^{-} \to \tau^{+}\tau^{-} (n\gamma)$, we estimate an uncertainty on the theoretical error corresponding to the truncation of the 
perturbative Feynman series as a function of the soft-photon cut-off applied in the exponentiation using the technique described in \cite{Nugent:2022ayu}. We discuss how this estimation procedure of the
theoretical error associated with the truncation of the perturbative Feynman series can be extended to most experimentally measurable physical observables and the potential pitfalls of this strategy.
\end{abstract}
\end{@twocolumnfalse}
]

\renewcommand{\thefootnote}{\fnsymbol{footnote}}
\footnotetext[1]{Corresponding Author \\ \indent   \ \ {\it Email:} inugent.physics@outlook.com}
\renewcommand{\thefootnote}{\arabic{footnote}}
\section{Introduction}

The summation of the infra-red divergences \cite{Bloch:1937} in the $e^{+}e^{-} \to \mu^{+}\mu^{-} (n\gamma)$  and $e^{+}e^{-} \to \tau^{+}\tau^{-} (n\gamma)$ processes are subtracted through the 
Yennie-Frautschi-Suura Exponentiation procedure \cite{Yennie:1961} in the {\tt ee$\in$MC} generator \cite{Nugent:2022ayu}. Within this formalism, all terms of the perturbative Feynman series related to the
QED processes  $e^{+}e^{-} \to \mu^{+}\mu^{-} (n\gamma)$  and $e^{+}e^{-} \to \tau^{+}\tau^{-} (n\gamma)$ contribute to their respective differential cross-section. Consequently, the largest theoretical error
associated with the predictions for the latter QED processes in {\tt ee$\in$MC} is due to the truncation of the perturbative Feynman series. The most significant of the neglected higher order terms in
\cite{Nugent:2022ayu} are related to the Initial-State and Final-State radiations. In the following sections, extension of the {\tt ee$\in$MC} generator \cite{Nugent:2022ayu} to take into account the leading
Initial/Final-State radiative diagrams with the Initial/Final interference at NLO, NNLO in conjunction with the leading Initial-State radiative diagrams at NNNLO will be examined. 
A generalized prescription for determining the optimal soft-photon cut-off and the corresponding theoretical uncertainty for the truncation of the perturbative Feynman series 
within the Yennie-Frautschi-Suura Exponentiation Formalism is also elucidated.         

\section{Exponentiation and Higher Order Radiative Emissions in {\tt ee$\in$MC} }

The total differentiable cross-section in terms of Initial and Final-State Yennie-Frautschi-Suura Exponentiation Form-Factors\footnote{The cross-terms for the  
Yennie-Frautschi-Suura Exponentiation Form-Factors are not included following the discussion in \cite{Yennie:1961}. Other generators have included \cite{kk2f} these terms.}, 
$Y_{f}(Q_{f}^{2})$ and $Y_{f}(Q_{f}^{2})$ respectively, may be written as

\begin{equation}
\resizebox{0.375\textwidth}{!}{$
d\sigma= \frac{\sum_{n=0}^{\infty} Y_{i}(Q_{i}^{2})Y_{f}(Q_{f}^{2})|\sum_{k=1}^{\infty}\bar{{\mathcal M}}_{n}^{k}|^{2} dPS_{n}}{4(|\vec{P}_{e^{-}}|E_{e^{+}}+E_{e^{-}}|\vec{P}_{e^{+}}|)}
$}
\label{eq:YSR}
\end{equation}

\begin{figure*}[tbp]
  \begin{center}
    \resizebox{520pt}{185pt}{
      \includegraphics{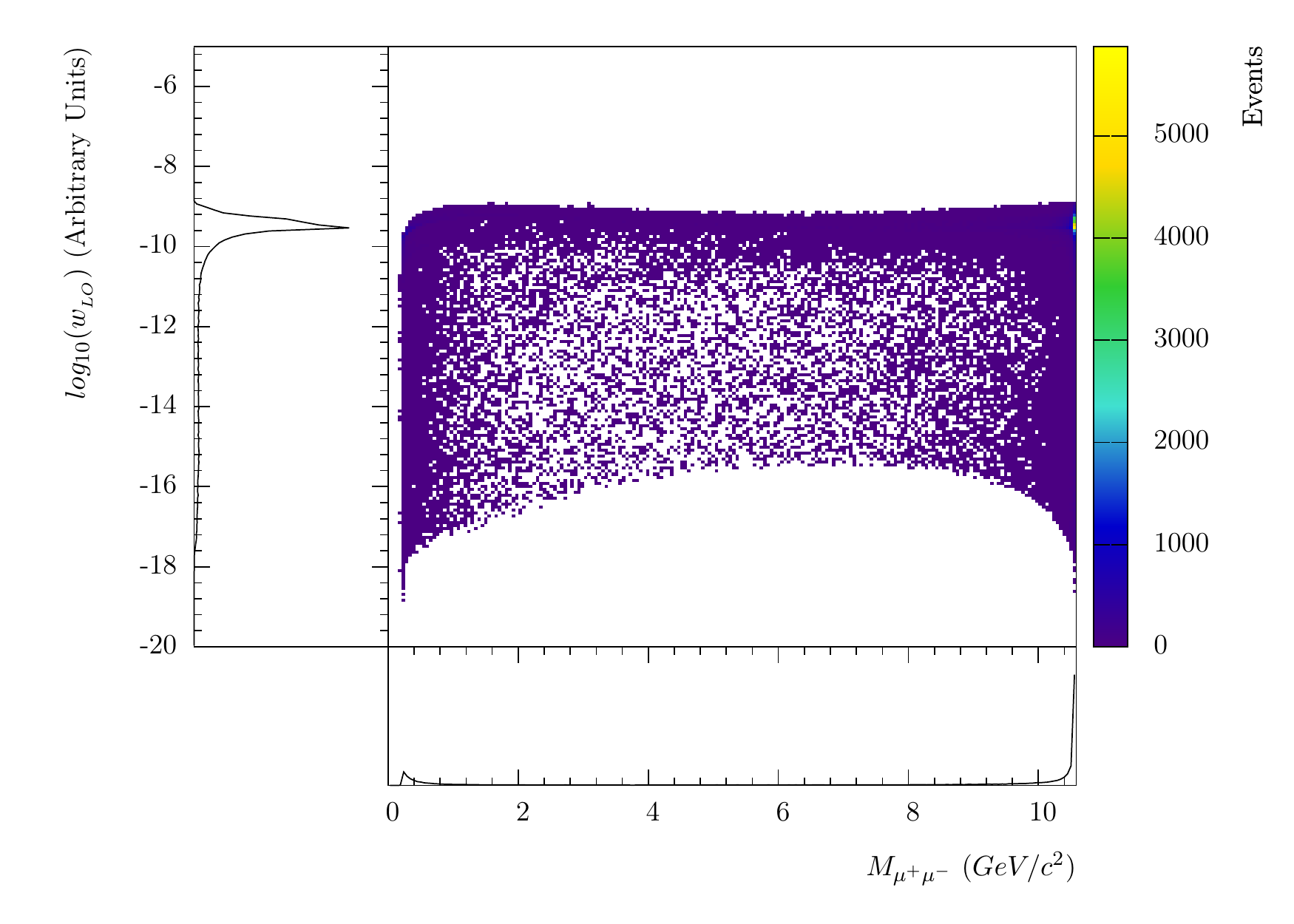}
      \includegraphics{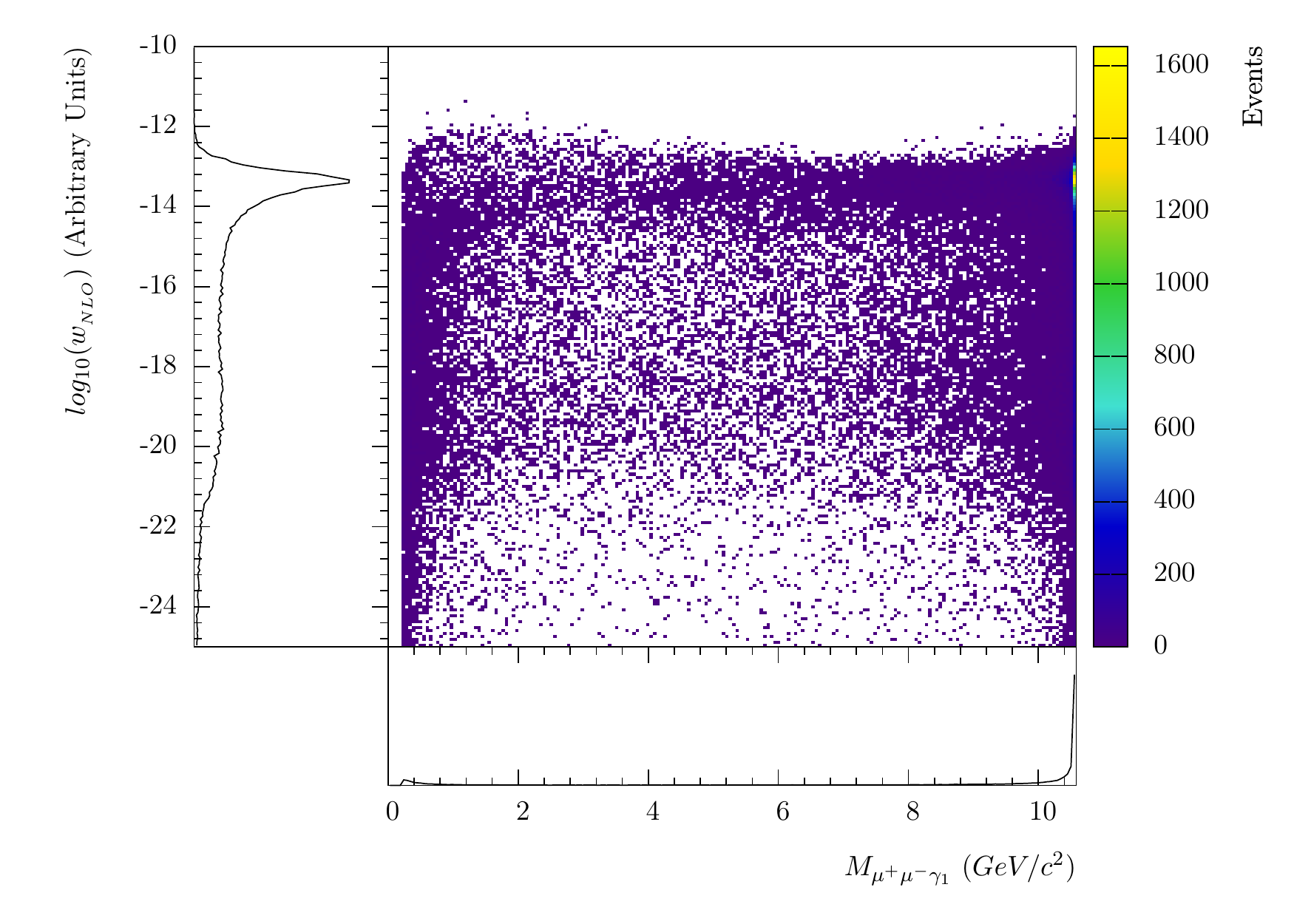}
    }
    \resizebox{520pt}{185pt}{
      \includegraphics{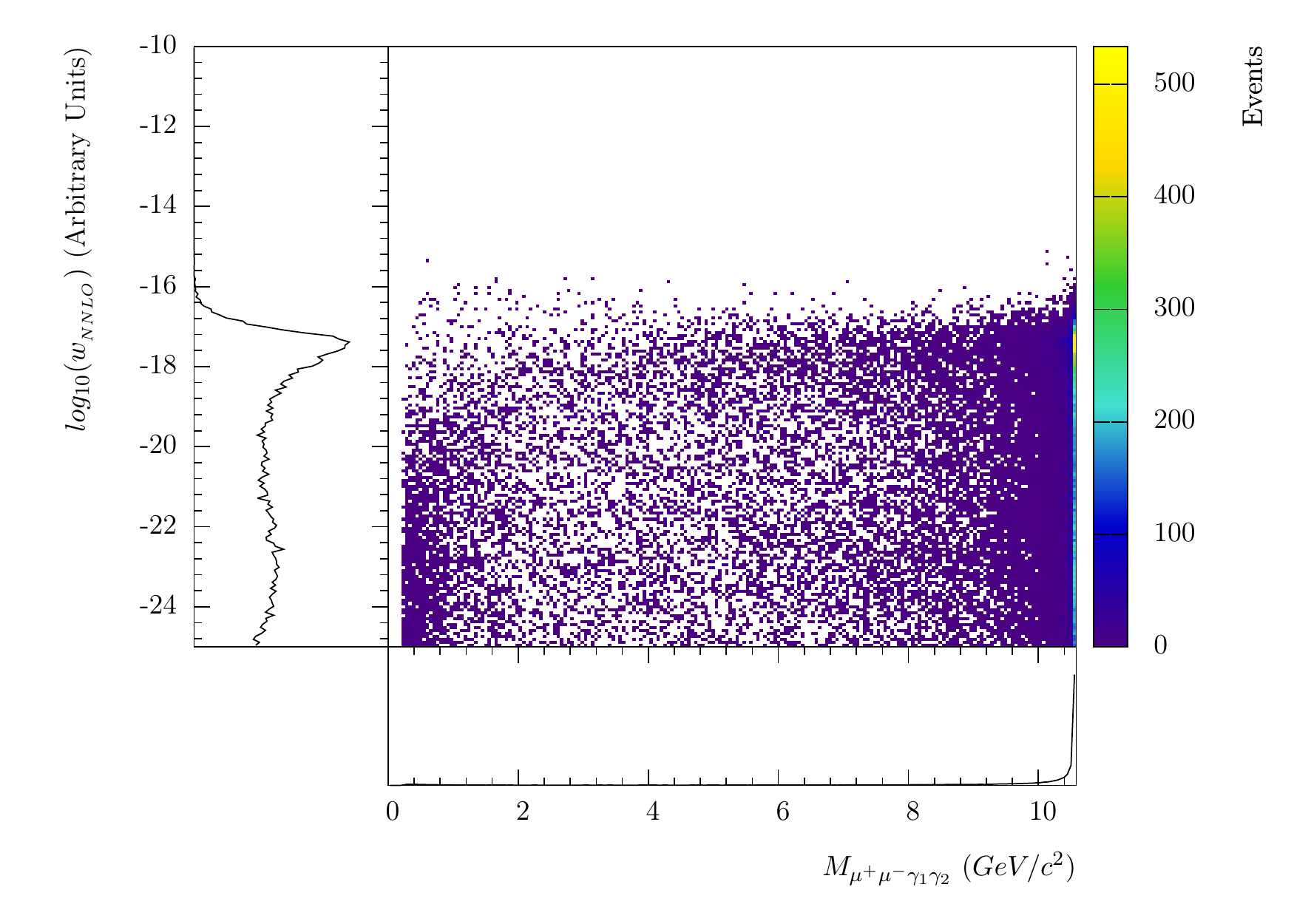}
      \includegraphics{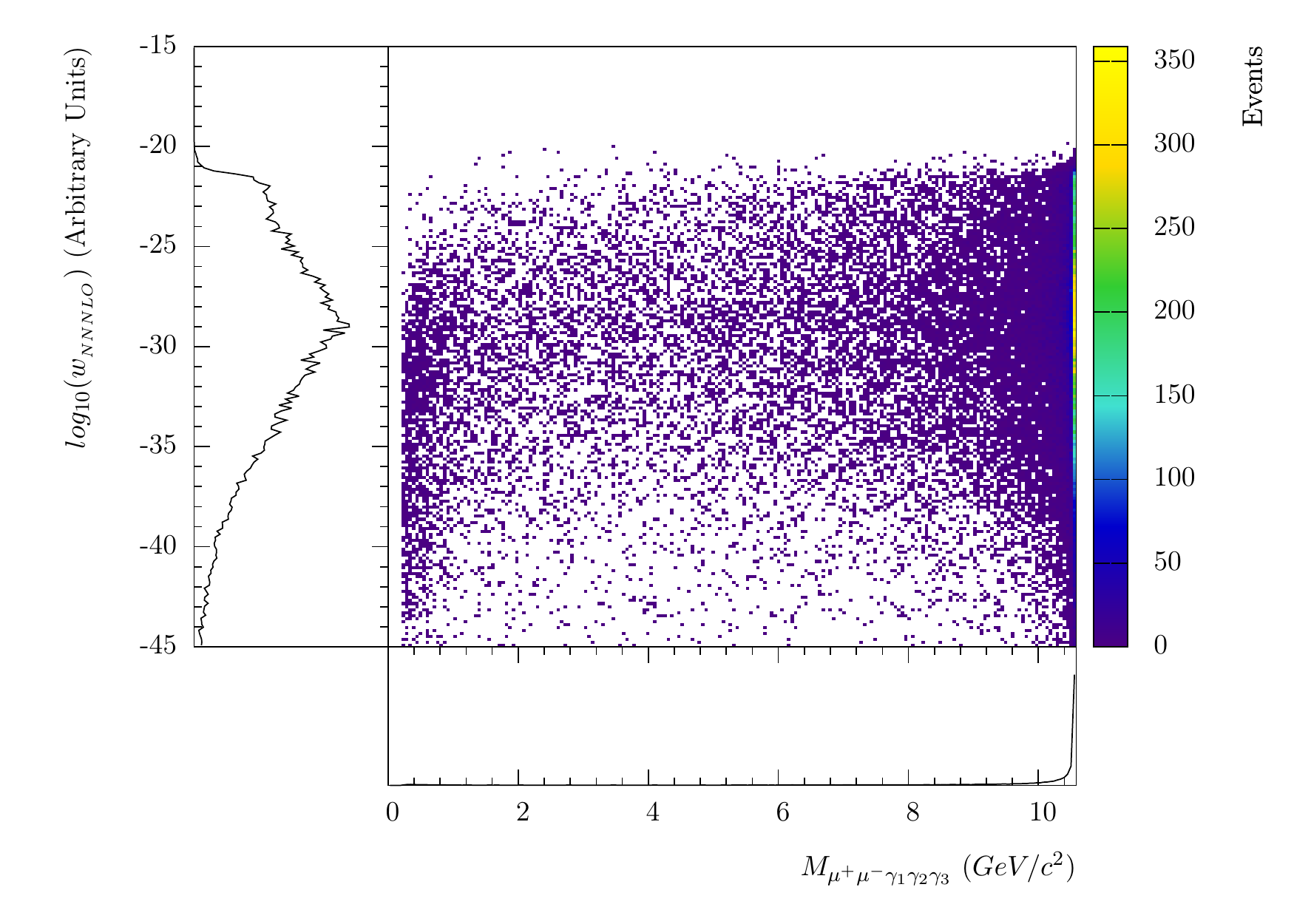}
    }
    \caption{The raw phase-space and QED weight distribution before applying the Jacobian corrections as a function of $M_{\mu^{+}\mu^{-}}$ invariant mass for LO  $\bar{{\mathcal M}}_{n=1}^{k=1}$  
      (upper-left), $M_{\mu^{+}\mu^{-}\gamma_{1}}$ invariant mass  for NLO $\bar{{\mathcal M}}_{n=2}^{k=1}$ (upper-right),
      $M_{\mu^{+}\mu^{-}\gamma_{1}\gamma_{2}}$ invariant mass for NNLO $\bar{{\mathcal M}}_{n=3}^{k=1}$ (lower-left) 
      and  $M_{\mu^{+}\mu^{-}\gamma_{1}\gamma_{2}\gamma_{3}}$ invariant mass for NNNLO $\bar{{\mathcal M}}_{n=4}^{k=1}$ (lower-right). The indices on the photons are related to the internal ordering of the 
      phase-space generator.  \label{fig:Weights}}
  \end{center}
\end{figure*}

\noindent \cite{Nugent:2022ayu,Yennie:1961}. Within {\it ee$\in$MC} there are several implementations of the Yennie-Frautschi-Suura Exponentiation Form-Factors: 
the Yennie-Frautschi-Suura calculation \cite{Yennie:1961};
the {\tt KK2F} approximation  \cite{kk2f}; Sudakov Form-Factor approximation \cite{Peskin:1995ev}; and from a calculation \cite{Nugent:2022ayu} based on the full QED LO calculation \cite{Schwinger:1998}. 
In the latter implementation, the Coulomb potential is subtracted from the QED LO calculation before computing the Exponentiation Form-Factors 
and applied as a separate correction factor \cite{Nugent:2022ayu,Peskin:1995ev,Smith1994117} to ensure that the Exponential Form-Factors satisfy the conditions of a multiplicative subtraction, 
$[0,1]$\cite{Peskin:1995ev}. 
For the remainder of this paper, we will restrict ourselves to the Yennie-Frautschi-Suura Exponentiation Form-Factors based on the full LO calculation from \cite{Schwinger:1998}, 
since it is the most precise calculation.
In Eq. \ref{eq:YSR}, the spin-average sum of the matrix element, $|\sum_{k=1}^{\infty}\bar{{\mathcal M}}_{n}^{k}|^{2}$, contains the sum of all diagrams with $n$  
hard-photon emissions and $k$ internal-photon lines. For each trial in the simulation, 
the matrix element is computed directly from the Feynman calculus using an object orientated formalism for the spinor and tensor algebra. Renormalization is implicitly included through the running of the 
QED coupling constant by means of Wards Identity \cite{Mandl:1985bg,Jegerlehner,Sturm_2013}\footnote{The limiting theoretical uncertainty in the running of the electromagnetic coupling, $\alpha_{EM}$,
comes from the comparison of the various schemes for extracting the hadronic vacuum polarization contributions from the experimental measurements \cite{Nugent:2022ayu}. The uncertainty 
associated with neglecting the contribution from higher order terms in the running of $\alpha_{EM}$ are substantially smaller. Therefore, it is reasonable to consider the higher order terms 
in the running of $\alpha_{EM}$ as an uncertainty on the prediction, thus allowing for the mis-match between the orders included renormalization and the QED calculations. }. 
The differential phase-space, $dPS_{n}$, is computed using a modified version of the mass-transfer algorithm from \cite{Byckling:1969} in which importance sampling \cite{Lopes:2006,Gelman:2014} 
is embedded to optimize the efficiency for a given physics model. 
The differential phase-space in Eq. \ref{eq:YSR} explicitly does not include the soft-photon cut-off since for higher orders it is applied to each individual vertex in the Feynman Diagrams.
The soft-photon cut-off is defined in terms of the Lorentz Invariant quantity $\delta M=M^{\prime}-M$ based on the Lorentz invariance, gauge invariance and renormalization invariance 
properties of the generalized Yennie-Frautschi-Suura Exponentiation Form-Factor \cite{Nugent:2022ayu}. For consistency with the conventional soft-photon cut-off energy, $E_{l}$, $\delta M$ is 
defined in \cite{Nugent:2022ayu} within the 1 photon limit. In \cite{Nugent:2022ayu},  $e^{+}e^{-} \to \mu^{+}\mu^{-} (n\gamma)$  and $e^{+}e^{-} \to \tau^{+}\tau^{-} (n\gamma)$ production was
simulated using both the Born ($\bar{{\mathcal M}}_{n=0}^{k=1}$) and LO diagrams ($\bar{{\mathcal M}}_{n=1}^{k=1}$), where the radiative NLO diagrams ($\bar{{\mathcal M}}_{n=2}^{k=1}$) were 
determined using weighted events. The phase-space generator has now been upgraded to a more generic algorithm which simulates $n$ hard-photons in all possible combinations corresponding to Initial-State and
Final-State photon geometries extending the embedded importance sampling and the mass-transfer phase-space algorithm in \cite{Nugent:2022ayu}. Examples of the weight distribution after applying the
importance sampling can be seen in Figure \ref{fig:Weights}. For $n\leq 4$, the phase-space run time is 
negligible compared to the physics model computation.  This has enabled the inclusion of all radiative emission Feynman diagrams associated with NLO 
($\bar{{\mathcal M}}_{n=2}^{k=1}$), NNLO ($\bar{{\mathcal M}}_{n=3}^{k=1}$) along with the NNNLO ($\bar{{\mathcal M}}_{n=4}^{k=1}$) radiative emission diagrams associated with Initial-State radiation.    

\section{Impact of Higher Order Radiative Emissions on the Cross-section and Physical Observables }

\begin{figure*}[tbp]
  \begin{center}
    \resizebox{520pt}{185pt}{
      \includegraphics{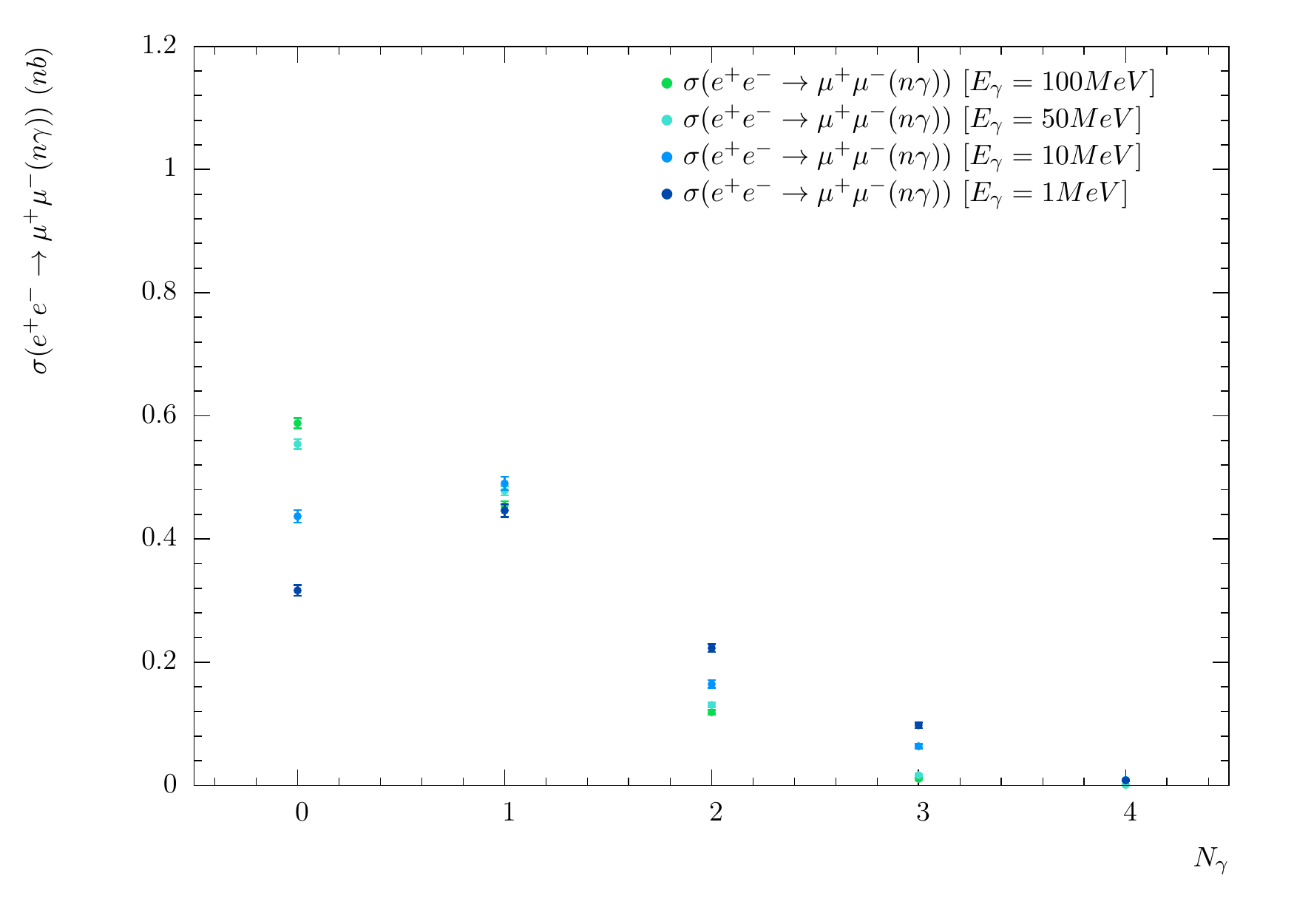}
      \includegraphics{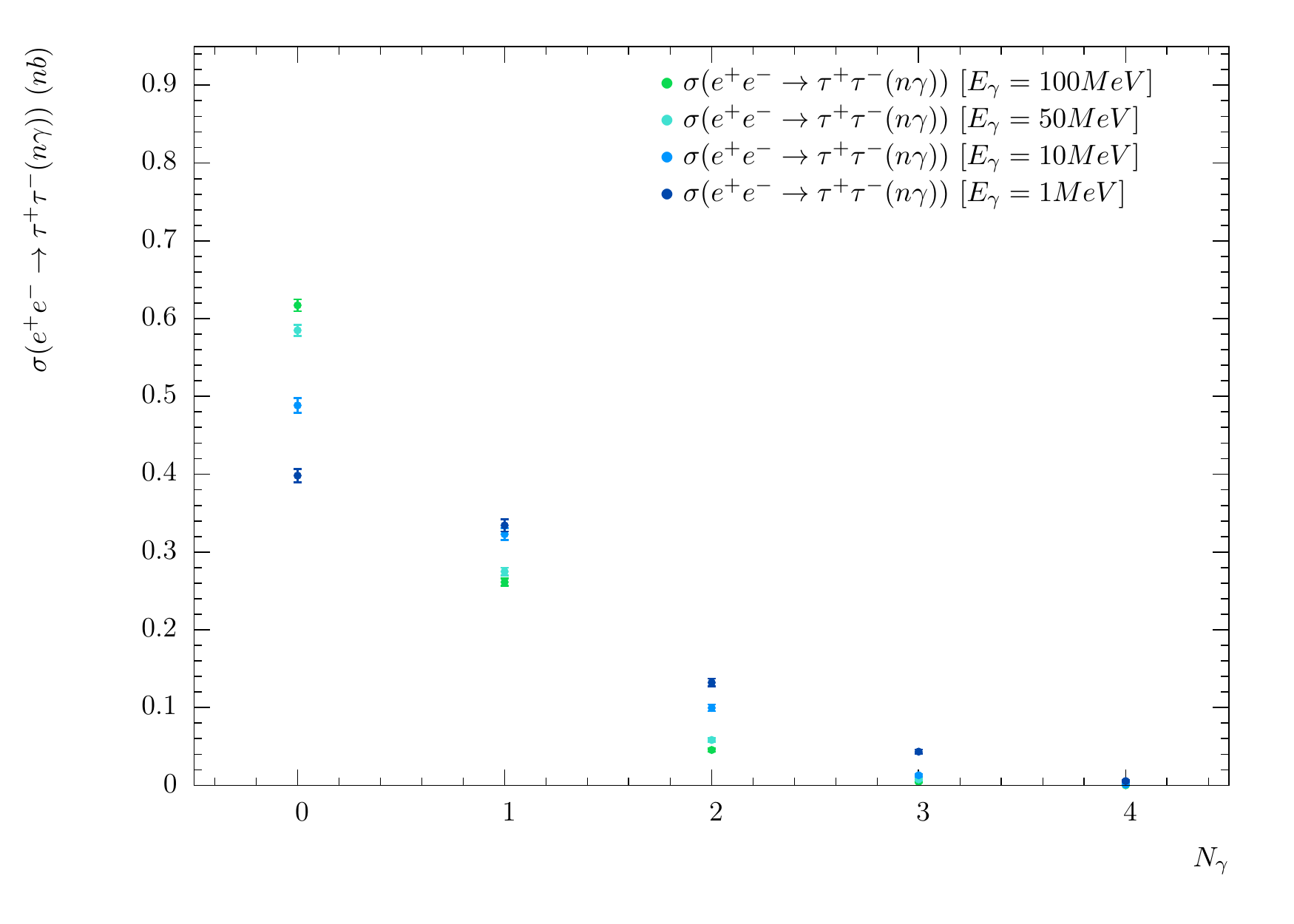}
    }
    \caption{The cross-section as a function of photon multiplicity for a soft-photon cut-off of $100MeV$, $50MeV$, $10MeV$ and $1MeV$ for the $e^{+}e^{-} \to \mu^{+}\mu^{-} (n\gamma)$ (left)  
      and $e^{+}e^{-} \to \tau^{+}\tau^{-} (n\gamma)$ (right) processes. 
      The error bar represents the statistical uncertainty on the number of events simulated for each of the individual photon cross-sections.  \label{fig:NGamma}}
  \end{center}
\end{figure*}

Within the construct of Yennie-Frautschi-Suura Exponentiation, the number of hard-photons emitted above a given energy are physical observables and are infra-red safe
after requiring the soft-photon cut-off energy, $E_{l}$, and minimum detectable photon energy of the given experiment, $E_{min}$, satisfies the condition $E_{min}\ge E_{l}$  
when the experiment is in the center-of mass frame
\cite{Yennie:1961,Peskin:1995ev}\footnote{In the literature, the soft-photon cut-off is most commonly evaluated in terms of the energy \cite{Yennie:1961,kk2f,Peskin:1995ev}. However, in
experiments where the laboratory and center-of-mass frame do not coincide the condition may be generalized to $E_{min}\gg \delta M\times c^{2}$, where $c=1$ in natural units.}.
Figure \ref{fig:NGamma} presents the $e^{+}e^{-} \to \mu^{+}\mu^{-} (n\gamma)$  and $e^{+}e^{-} \to \tau^{+}\tau^{-} (n\gamma)$ cross-section as a function of the number of
hard-photons for several soft-photon cut-offs.
The soft-photon cut-off energies correspond to the typical minimum measurable photon energy, $E_{min}$, in the
laboratory frame at $e^{+}e^{-}$ experiments, such as BELLE and BABAR, $50-100MeV$ \cite{Bevan_2014}. Although, for these experiments, the soft-photon cut-off energy does not directly correspond to the 
$E_{min}$ due to the boost between the center-of-mass reference frame and laboratory reference frame, it demonstrates that there is a non-negligible contribution from $n=2,3$ and $4$ hard-photons
for $e^{+}e^{-} \to \mu^{+}\mu^{-} (n\gamma)$  and $e^{+}e^{-} \to \tau^{+}\tau^{-} (n\gamma)$ interactions at these experiments.  
The simulated photon distributions can be comparable with the expected Poisson distribution from the classical limit \cite{Peskin:1995ev} for photon energies within a given energy.
The total cross-section, as a function of soft-photon cut-off, is presented in Figure \ref{fig:csvscutoff} for both $e^{+}e^{-} \to \mu^{+}\mu^{-} (n\gamma)$  
and $e^{+}e^{-} \to \tau^{+}\tau^{-} (n\gamma)$ processes. If all orders of the perturbative Feynman series were included in the calculation the cross-section would be
infra-red safe and therefore independent of the soft-photon cut-off. Thus, a dependency of the cross-section on the soft-photon cut-off is a consequence of the truncation of the perturbative Feynman series.
More specifically, the magnitude of the higher order terms which are truncated increase as the soft-photon cut-off is decreased\footnote{This interpretation is dependant on the positive definite 
multiplicative subtraction of the infra-red divergencies \cite{Yennie:1961}.}.
This increasing significance of the higher order terms with decreasing soft-photon cut-off is analogous to the LO in the effective mass scheme for infra-red subtraction developed by Feynman 
\cite{Griffiths:1987tj}.
In terms of choosing the optimal $\delta M$ and assessing the theoretical uncertainty and precision of the predictions 
from {\it ee$\in$MC} related to the soft-photon cut-off within the Yennie-Frautschi-Suura Exponentiation Formalism there are three issues which must be taken into account:

\begin{enumerate}
\item Firstly, there is a truncation error associated with neglecting higher order terms in the perturbative Feynman series associated with a given soft-photon cut-off. Within the 
Yennie-Frautschi-Suura Exponentiation Formalism, the perturbative Feynman series is infinite and for physical observables converges asymptotically to an infra-red safe value as the number of terms 
included goes to infinity. However, for the small coupling constant in QED, it generally converges rapidly enough that only the first few 
terms are required. The higher order terms become more significant as the soft-photon cut-off is reduced. This causes a dependency of the simulated observables, for example the cross-section,
on the soft-photon cut-off.  Exploiting this, the deviation in the prediction of an infra-red safe observable computed with a given soft-photon cut-off compared to the value at a 
higher soft-photon cut-off where the contributions from higher order terms is negligible, can be used to estimate the theoretical uncertainty related to the truncations \cite{Nugent:2022ayu}. In 
general, the error resulting from the 
truncation of the perturbative Feynman series can be calculated for any given physical effect or observable if and only if one or more of the dominate Feynman diagrams which cause the effect
are included in the simulation. Particular examples of this in {\it ee$\in$MC} include neglecting the $Z^{0}$ boson, for the interference between internal photon lines 
from diagrams with 2 or more internal photon lines. For an infra-red safe observable, this method is expected to be more reliable for estimating the truncation error than from the  magnitude of the next most 
significant higher order terms 
since it includes all higher order terms and does not rely on estimating which terms are most important \footnote{From Figure \ref{fig:NGamma}, the contribution of the 4 hard-photons 
Initial-State radiation at NNNLO can be used as an estimate of the truncation error from radiative diagrams. At $\delta M=1MeV$, the 4 hard-photon Initial-State
radiation NNNLO contribution to the  $e^{+}e^{-} \to \mu^{+}\mu^{-} (n\gamma)$ cross-section is $\sim 0.006-0.008nb$. The full truncation error on the NNLO  $e^{+}e^{-} \to \mu^{+}\mu^{-} (n\gamma)$ 
cross-section must necessarily be greater than the contribution from the 4 hard-photon Initial-State
radiation NNNLO contribution.}.

\item Secondly, for a meaningful and internally consistent quantum mechanical interpretation of the summation over the soft-photons, 
it is essential that the energy from the soft-photons in the laboratory reference frame is less than the minimal experimentally 
measurable photon energy, $E_{min}$ \cite{Peskin:1995ev}. This is the criteria that ensures the number of hard-photons is an infra-red safe observable. This implies that $\delta M$ must 
necessarily be chosen to be sufficiently small so that in the laboratory frame the number of photons 
with $E_{l}>E_{min}$ from vertices with $M^{\prime}-M<\delta M$ is negligible\footnote{Constructing the soft-photon cut-off in terms of $\delta M$ has the advantage, that the procedure is 
the same in the B-Factories which have a boost relative to the center-of-mass frame and experiments in which the laboratory reference frame and center-of-mass reference frame coincide.}. 
For regions of $\delta M$ where the truncation error is negligible, this condition can be confirmed by comparing the photon multiplicity distribution for events generated 
with two distinct soft-photon cut-offs, both of which are expected to satisfy the condition, 
after requiring a minimum experimental energy for the measured photon within the laboratory reference frame\footnote{If a given 
soft-photon cut-off, $\delta M_{A}$, satisfies this criteria, all cut-off values equal to or below $\delta M_{A}$ ( $\delta M_{A}\ge\delta M$) also satisfy this criteria independent of the truncation error.}. 
In this case, a statistically significant deviation between the samples is a clear indication that the condition is not satisfied and a smaller value of $\delta M$ is required. Alternatively, 
if the deviation is sufficiently small, a theoretical uncertainty may be applied.  

\item In addition to the theoretical uncertainty related to the choice of the soft-photon cut-off $\delta M$, there is also a discrepancy between the theoretical predictions
and the expectation in the data related to a discontinuity of the mass spectrum of the final state lepton pair caused by the soft-photon cut-off, $\delta M$. 
More specifically, the soft-photon cut-off $\delta M$ results in an upper boundary on the lepton pair mass in radiative LO and higher order radiative terms 
which only tends to the Born mass of the lepton pair in the limit${}_{\delta M \to 0}$. When $\delta M$ is comparable to or larger than the experimental resolution of the outgoing lepton pair mass, 
this effect can cause a
non-negligible bias in the lepton pair mass spectrum near the Born lepton pair mass peak.  
\end{enumerate}

\section{Conclusions}

\begin{figure*}[tbp]
  \begin{center}
    \resizebox{260pt}{185pt}{
      \includegraphics{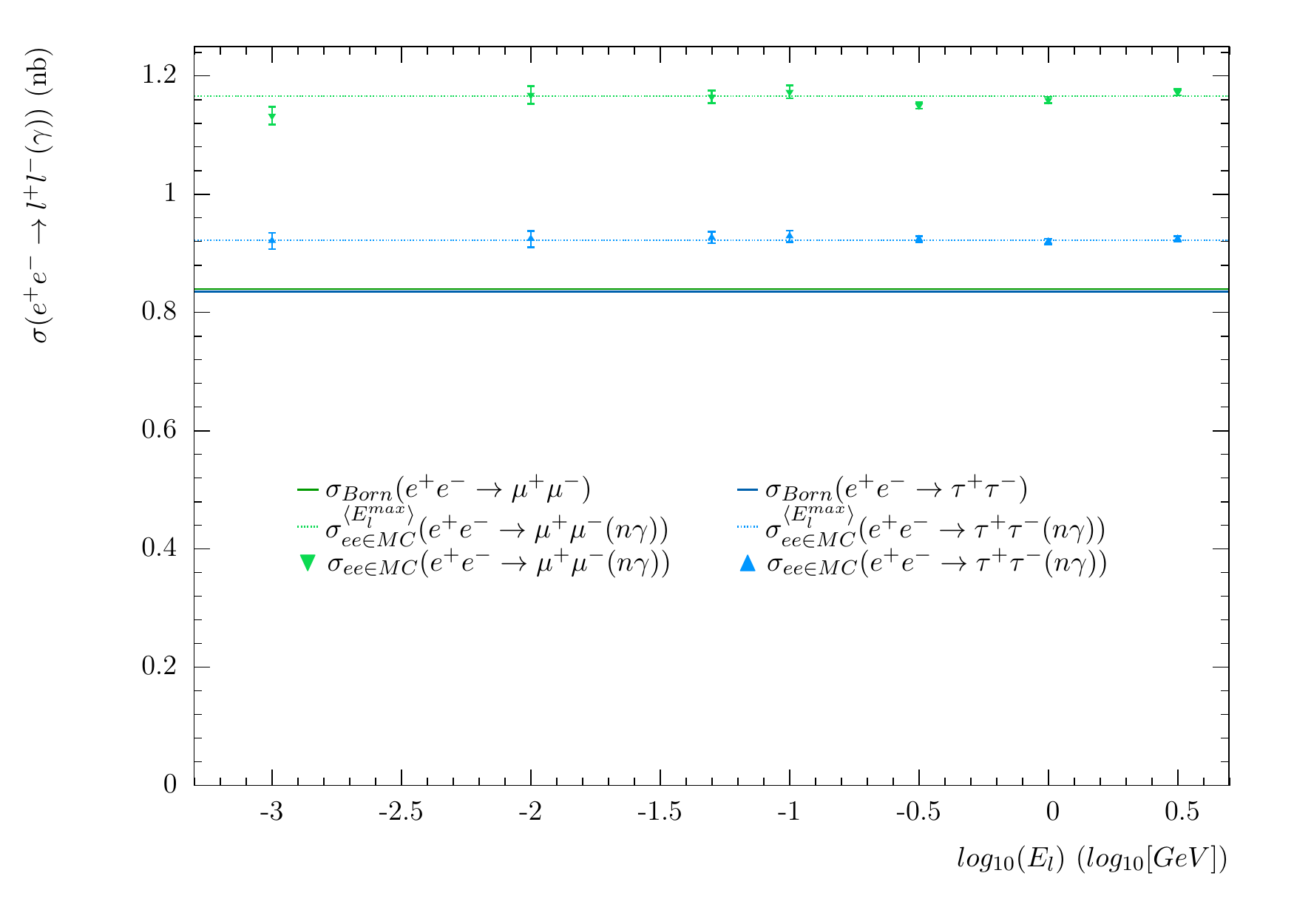}
    }
    \caption{The total cross-section as a function of soft-photon cut-off which is used to demonstrate the magnitude of the truncation error at a given soft-photon cut-off for the
      $e^{+}e^{-} \to \mu^{+}\mu^{-} (n\gamma)$  and $e^{+}e^{-} \to \tau^{+}\tau^{-} (n\gamma)$ processes. The error bars
      represent the statistical uncertainty on the number of events simulated. $\sigma_{ee\in MC}^{\langle E_{l}^{max}\rangle}$, the average of the cross-sections for the upper most $E_{l}$ which 
      were shown to have converged by NLO in \cite{Nugent:2022ayu},
      are overlayed as dotted lines to illustrate the expected converged cross-sections. Any statistically significant deviation from this is an indication of a 
      non-negligible truncation error. \label{fig:csvscutoff} }
  \end{center}
\end{figure*}

Higher order radiative diagrams corresponding to Initial/Final-State radiative emissions, including interference, for NLO ($\bar{{\mathcal M}}_{n=2}^{k=1}$), NNLO ($\bar{{\mathcal M}}_{n=3}^{k=1}$) 
in conjunction with the Initial-State NNNLO ($\bar{{\mathcal M}}_{n=4}^{k=1}$) have been implemented in {\it ee$\in$MC} for the $e^{+}e^{-} \to \mu^{+}\mu^{-} (n\gamma)$ and 
$e^{+}e^{-} \to \tau^{+}\tau^{-} (n\gamma)$ interactions. This produces a significant improvement in the convergence of the total cross-section as a function of soft-photon cut-off, 
as seen in Figure \ref{fig:csvscutoff}. The criteria for determining the soft-photon cut-off, $\delta M$, within the Yennie-Frautschi-Suura Exponentiation Formalism are defined 
and the uncertainties corresponding to each of these conditions are discussed. This includes a generalized prescription for the truncation uncertainty which is suitable for most experimental observables
for which the corresponding physics is contained within the Feynman diagrams being simulated.  From Figure \ref{fig:csvscutoff}, the $e^{+}e^{-} \to \tau^{+}\tau^{-} (n\gamma)$ cross-section 
is statistically consistent with a negligible truncation error on the perturbative Feynman series for soft-photon cut-offs down to $\sim1MeV$. Thus 
the theoretical uncertainty on the truncation error is soley due to the statistical uncertainty on 
the number of simulated events.  In contrast, the $e^{+}e^{-} \to \mu^{+}\mu^{-} (n\gamma)$ cross-section has a slight statistical deviation which indicates that for a
soft-photon cut-off of $1MeV$ the cross-section is beginning to decrease. In this case, the statistical uncertainty and the deviation should be added in quadrature ($\pm3.2\%$). For a soft-photon cut-off of $1MeV$ in the
 $e^{+}e^{-} \to \mu^{+}\mu^{-} (n\gamma)$ interaction, 
the NNNLO Initial-State radiative emission contribution is $\sim 0.006-0.008nb$ (or $\sim0.7\%$)\footnote{This estimate is based on the both Figure \ref{fig:csvscutoff} and 
from simulations using weighted events.}. 
This is a minimum uncertainty when neglecting radiative emissions from NNNLO and higher. Although additional statistics are required to estimate a 
competitive uncertainty on the truncation error, the inclusion of the $\bar{{\mathcal M}}_{n=2}^{k=1}$, $\bar{{\mathcal M}}_{n=3}^{k=1}$, and $\bar{{\mathcal M}}_{n=4}^{k=1}$ terms in {\it ee$\in$MC}
have significantly reduced the truncation error in the theoretical predictions.   

\section*{Acknowledgement}
GCC Version 4.8.5 was used for compilation and the plots are generated using the external program GNUPlot  \cite{gnuplot4.2}.

\footnotesize
\bibliography{paper}
\normalsize

\end{document}